\documentstyle[11pt]{article}
\begin{document}
\title{\sc How Can Strange Quark Matter Occur Deeply in the
Atmosphere?}
\vspace{-5mm}
\author{G.Wilk$^{1}$\thanks{e-mail: wilk@fuw.edu.pl} 
and Z.W\l odarczyk$^{2}$\thanks{e-mail: wsp-fiz@srv1.tu.kielce.pl}} 
\date{
\begin{center}
\small
$^1${\it Soltan Institute for Nuclear Studies, Warsaw, Poland,}\\
$^2${\it Institute of Physics, Pedagogical University, Kielce,
Poland} 
\end{center}
%\today}
}
\maketitle
%\newpage
\begin{abstract}
Motivated by some recent cosmic ray experiments we study the
properties of strange quark matter near flavour equilibrium. Using  
Fermi-gas model we argue that, contrary to some claims, the
geometrical radii of quark matter strangelets are not smaller but
rather comparable to those of ordinary nuclei. We propose therefore a
scenario of propagation of strangelets through the atmosphere
which still allows for their deep penetration.\\ 

PACS: 13.85.Tp~~~12.90.+b
\end{abstract}
\vspace{7mm}

\centerline{Preprint: {\bf SINS-PVIII-1996-1}
and {\bf IP-PUK-1996-5}}
\vspace{12mm}

%\newpage
In astrophysical literature one finds a number of phenomena that are
both puzzling and extremaly unusual or unique. They are interested
because they apparently do not belong to any known classes of
registered events. Some of them are regarded as a manifestation of
the so called {\it strange matter}. In particular it concerns such
events as: anomalous cosmic ray bursts from {\it Cygnus} $X-3$
\cite{BAYM}; extraordinary high luminosity gamma-ray bursts from the 
supernova remnant $N49$ in the Large Magellanic Cloud \cite{AFO}; the
so called {\it Centauro} events \cite{LATTES} (characterised by
anomalous composition of secondary particles with almost no neutral
pions present). In this last case, if it would really be caused by
some kind of {\it strange quark matter} (SQM) called {\it
strangelets}, their observation at the atmospheric depth as large as
$\sim 500$ g/cm$^2$ would require unusual penetrability of these
lumps of SQM \cite{BJORKEN} (i.e., their geometrical sizes or
interaction cross sections should be much smaller than the typical
nuclear size).\\   

Typical SQM consists of roughly equal number of up, down and strange
quarks and it has been found to be the true ground state of QCD
\cite{WITTEN,FARHI,BERGER}\footnote{It is therefore absolutely stable
at high mass numbers $A$ and because the energy per baryon in SQM
could be smaller than that in ordinary nuclear matter, it would be
more stable than the most tightly bound $^{56}$Fe nucleus.}. However,
any SQM produced at very early stage of the history of Universe would
have evaporated long time ago \cite{ALOCK}. On the other hand, there
are places where the SQM may still exists at present
\cite{WITTEN,ALOCK}. It is probably continously produced in neutron
stars with a superdense quark surface and in quarks stars with a thin
nucleon envelope \cite{HAENSEL} and collisions of such objects could
produce small lumps of SQM, strangelets with $10^2 < A < 10^6$,
permeating the Galaxy and possibly reaching also the Earth (i.e.,
{\it a priori} being detectable here).\\  

There are several reports suggesting direct candidates for SQM.
In particular, anomalous massive particles, which can be interpreted
as strangelets, have been apparently observed in three independent
cosmic ray (CR) experiments:
\begin{itemize}
\item[$(i)$] In counter experiment devoted to study primary CR nuclei
two anomalous events have been observed \cite{SAITO}. They are
consistent with values of charge $Z \simeq 14$ and of atomic numbers
$A \simeq 350$ and $\simeq 450$ and cannot be accounted for by the
conventional background. Such values of $Z$ and $A$ are fully
consistent with the theoretical estimation for ratio $Z/A$ in SQM 
\cite{KASUYA}.
\item[$(ii)$] The so called Price's event \cite{PRICE} with $Z \simeq
46$ and $A > 1000$, regarded previously as possible candidate for
magnetic monopole, turned out to be fully consistent with the above 
ratio $Z/A$ for SQM \cite{S}.
\item[$(iii)$] The so called Exotic Track with $Z \simeq 20$ and $A
\simeq 460$ has been reported in \cite{ICH}. The name comes from the
fact that although it was observed in emulsion chamber exposed to CR
on balloon at the atmospheric depth of only $11.7$ g/cm$^2$ its
arrival zenith angle of $87.4$ deg means that the projectile causing
that event traversed $\sim 200$ g/cm$^2$ of atmosphere (in contract
to events $(i)$ and $(ii)$ where the corresponding depths were of the
order of $5 - 15$ g/cm$^2$ of atmosphere only). 
\end{itemize}
The Exotic Track event motivated the (balloon born emulsion chamber)
JACEE \cite{JACEE} and Concorde aircraft \cite{CAP} experiments to
search for strangelets with such long mean free paths. In fact,
authors of \cite{JACEE,CAP} suggest that the interaction mean free
path for strangelets in atmosphere is of the order of $\lambda_{\rm
S} = 124$ g/cm$^2$ for $A = 100$ decreasing to $\lambda_{\rm S} = 59$
g/cm$^2$ only for $A = 1000$. These values are surprisingly close 
to  that for protons at comparable energies ($\lambda_{proton} =
60~-70~$  g/cm$^2$) and are much bigger than that for a normal
nucleus ($\lambda_{nucleus} \simeq 3.8$ g/cm$^2$ for $A = 100$). This
means that (according to \cite{JACEE,CAP}) strangelets should have
geometrical radii much smaller than those of the ordinary nuclei or,
correspondingly, much smaller interaction cross sections (this would
also agree with the SQM interpretation of Centauro  events mentioned
before)\footnote{So far the results are negative: no evidence for
strangelets with $Z > 26 $ that survived passing of $\sim 100$
g/cm$^2$ of atmosphere was found \cite{JACEE,CAP}.}.\\ 

In this Letter we would like to show that, contrary to the above
expectations, geometrical radii of strangelets are most probably
comparable to those of the ordinary nuclei. Nevertheless, we shall
also show that it is still possible to expect some strangelets
occuring deeply in the atmosphere as advocated by
\cite{ICH,JACEE,CAP}.\\ 

Let us consider a lump of SQM visualised after \cite{FARHI,BERGER}
as Fermi gas of up, down and strange quarks, with total mass number
$A$ and confined in a spherical volume $V \sim A$. Its radius is
given by 
\begin{equation}
R\, =\, r_0\,A^{1/3} , \label{eq:R}
\end{equation}
where the rescaled radius $r_0$ is determined by the number density
of the strange matter, $n = A/V$ \cite{BERGER}. This in turn is given
by the sum of densities of each quark species under consideration,
$n\, =\, \frac{1}{3}\, \left( \, n_u\, +\, n_d\, +\, n_s\, \right)$,
where
\begin{equation}
n_i \, =\, -\, \frac{\partial \Omega _i}{\partial \mu _i} \label{eq:OM}
\end{equation}
and thermodynamical potentials $\Omega_i(m_i,\mu_i)$ are related to
chemical potentials $\mu_i$. Because chemical potentials of interest
here are of the order of $\mu \sim 300$ MeV, one can neglect the
(current) masses of up and down quarks and leave only the mass of the
strange quark, which we shall denote by $m$. Taking into account the
QCD ${\cal O}(\alpha_c)$ corrections to the properties of SQM in
calculating the respective thermodynamical potentials
$\Omega_i(m_i,\mu_i,\alpha_c)$ (renormalizing them at $m_N/3 = 313$
MeV) \cite{FARHI}, the rescaled radius is given by 
\begin{equation}
r_0\, =\, \left[\, \frac{3\, \pi}
                        {2\, \left(1 - \frac{2\alpha_c}{\pi}\right) 
                        \, \left( \mu^3 + m^3 \right) }\, \right]
           ^{1/3} . \label{eq:r0}
\end{equation}
In Fig. $1a$ we show its dependence on the ratio of the strange quark
mass to its chemical potential, $m/\mu$ for the case of $\alpha_c=0$
(i.e., ignoring one-gluon exchanges inside the Fermi gas
\cite{BERGER}). For the values commonly accepted for SQM (like $m
\simeq 150$ MeV and $\mu \simeq 300$ MeV) \cite{WITTEN,FARHI,BERGER},
the values of $r_0$ of the strangelets are comparable with that for
the ordinary nuclear matter (being only a bit smaller with difference
not exceeding $10$\% - $20$\%). Fig. $1b$ summarizes the dependence
of the $r_0$ on the QCD coupling constant $\alpha_c$. It turns out
that in this case the  chemical equilibrium shifts towards bigger
number of strange quarks without, however, significantly influencing
the number density. As one can see the QCD corrections lead to the
slight increase of $r_0$ (not exceeding $30$\%).\\ 

Fig. $1$ shows therefore that the expected decrease of the radius of
strangelet is nowhere as dramatic as it has been estimated in Refs.
\cite{JACEE,CAP}. It means that the expected geometrical cross
sections of SQM are, in fact, not much smaller than those for normal
nuclei, in any case not enough to explain alone the occurences of
anomalous events detected deeply in the atmosphere. It does not mean,
however, that some SQM cannot be registered there and in what follows
we shall propose a simple (speculative but plausible) scenario, which
can be summarized as follows: strangelets reaching so deeply into
atmosphere are formed in many successive interactions with air nuclei
of much heavier lumps of SQM entering our atmosphere. Assuming now
the simplest possible scenario, namely that after every such
collision strangelet of mass number $A_0$ becomes the new one with
$A_0 - A_{air}$ \footnote{Notice that incoming strangelet has $A_0 >>
A_{air}$ and that it is much more stable than the target air
nucleus. For simplicity we tacidly assume here that air nucleus
destroys totally the corresponding (equal to it) part of the incoming
lump of SQM, our estimation provides therefore a lower limit of what
should be expected in more detailed calculations.}, one obtains the
resultant mass number of strangelet registered at depth $h$, $A(h)$,
as a function of $h$ as shown in Fig. $2$. As one can see, bigger
initial strangelets (i.e., with higher mass number $A_0$) can
penetrate much more deeply into atmosphere untill $A(h)$ exceeds
critical $A_{crit}$, after which point they just evaporate by the
emission of neutrons.\\

Let us now explain our point in more detail. First of all let us note
that the practical measure of the stability of strangelet is the so
called separation energy $dE/dA$, i.e., energy, which is required to
remove a single baryon from a given strangelet. For example, if
$dE/dA > m_N$ strangelet can evaporate (from its surface)
neutrons\footnote{In principle strangelet is unstable if its energy
per nucleon (understood here as $3$ quarks), $E/A$, exceeds the mass
of a nuclear system. However, even if $E/A > m_N$ a strangelet would
not convert as a whole into nucleons in any finite time because such
process would be of extremaly high order in the weak coupling
constant.}. This energy depends, among other things, on the size of
the strangelet, which is usualy given in terms of its mass number $A$
\cite{BERGER}. There exists therefore some critical size given by a
critical value of $A = A_{crit}$ such that   for $A > A_{crit}$ there
will exist some strangelets that are absolutely stable against
neutron emission; it depends on the various choices of parameters and
vary from $A_{crit} = 300$ to $400$ \cite{FARHI,BERGER}. Suppose now
that the energy per baryon in strange matter is $\varepsilon = 919$
MeV \cite{FARHI} and that number densities corresponding to nuclear
matter are $n = (110~{\rm MeV})^3$. In such situation for $A\leq
1100,~E/A$ exceeds already $m_N$ but strangelet does not emit
neutrons yet and starts to do so only for $A\leq 320$, at which point
$dE/dA$ exceeds $m_N$ \cite{FARHI}. Below this limit strangelet
decays rapidly by evaporating neutrons. In view of these remarks it
is remarkable that all possible candidates for SQM have mass numbers
near or slightly exceeding $A_{crit}$, namely: $A = 350$ and $450$ in
\cite{SAITO}, $A = 460$ in \cite{ICH} and $A = 1000$ in \cite{S} and
is argued that Centauro event contains probably $\sim 200$ baryons
\cite{BJORKEN,LATTES}. Fig. $3$ shows atmospheric length traversed
after which the strangelet mass number $A$ becomes critical, $A =
A_{crit}$, starting from different mass numbers $A_0$ of the initial
strangelets. From it one can read off that strangelets which are
observed at depth $200$ g/cm$^2$ should originate from the
strangelets of mass number $A_0 = 900$ at the top of the atmosphere
whereas Centauro events observed at the mountain altitudes would
require original strangelet of $A_0 = 1800$. Unfortunately, the mass
distribution of incoming strangelets, $N(A_0)$, is not known.
Assuming that strange stars break up in the collisions in a manner
resembling the breaking of colliding nuclei (proceeding due to the 
phase transition close to the critical point), one could expect that
the corresponding mass distribution of strangelets should follow a
simple power law, like $A_0^{\gamma}$ \footnote{As a good candidate
for its statistical interpretation could serve percolation model of
Ref. \cite{NEMETH}.}. In such a picture power index $\gamma \geq 2$
can accomodate both the observed strangelets and the upper bounds for
intensities of those which survived traversing $\sim 100$ g/cm$^2$ of
atmosphere as provided by \cite{JACEE,CAP}.\\

We summarize by stating that most probable the geometrical cross
sections of strangelets are not dramatically different from those for
the ordinary nuclear matter and cannot therefore explain their
apparent very high penetrability through the atmosphere. Instead we
propose to interprete such a penetrability of SQM (already discovered
or to be yet observed) as indication of the existence of very heavy
lumps of SQM entering our atmosphere, which are then decreased in
size during their consecutive collisions with air nuclei (i.e., their
original mass number $A_0$ is reduced until $A = A_{crit}$) and
finally decay be the evaporation of neutrons.\\

\newpage
\begin{figure}[h]
\setlength{\unitlength}{1in}
\begin{picture}(3.021,5.5979)
\includegraphics{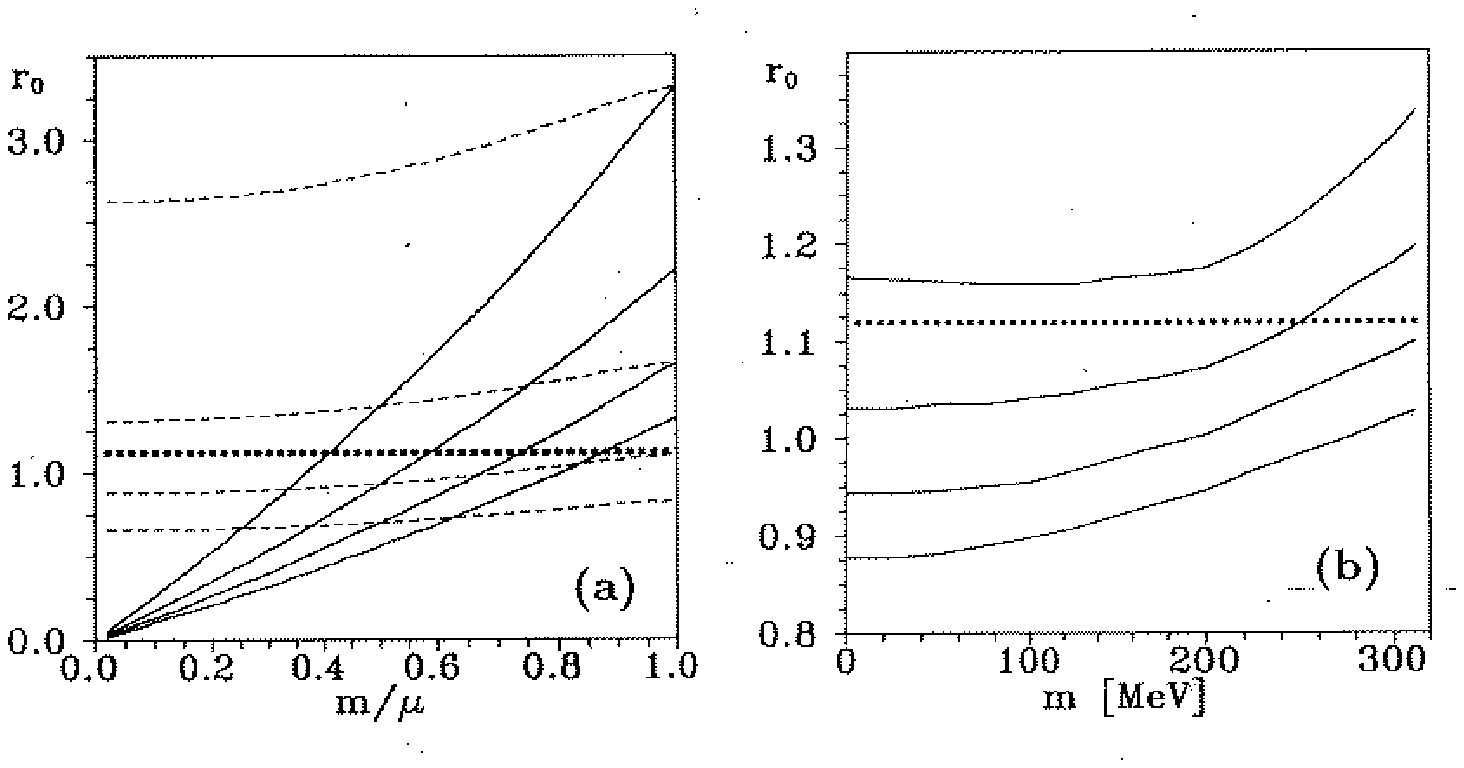}
\end{picture}
\caption{Dependence of the rescaled radius $r_0$ [fm] (cf.
eq.(3): $(a)$ on the ratio of the strange quark mass $m$ to
its chemical potential $\mu$, $m/\mu$ (solid lines, read from top to
bottom, correspond to fixed strange quark masses: $m =
100,~150,~200,~250$ MeV; dashed read from top to bottom correspond to
fixed chemical potentials: $\mu = 100,~ 200,~300,~400$ MeV); $(b)$ on
the strange quark mass $m$ for different values of the QCD coupling
constant: $\alpha_c = 0.9,~0.6,~0.3,~0.0$ (from top to bottom,
respectively). In both cases we show also for reference (by dotted
line) the $r_0 = 1.12$ fm corresponding to normal nuclear density
$\rho = 0.17$~\rm fm$^{-3}~=~(110$~\rm MeV$)^3$.  }
\end{figure}
\newpage
\begin{figure}[h]
\setlength{\unitlength}{1in}
\begin{picture}(4.867,5.000)
\includegraphics{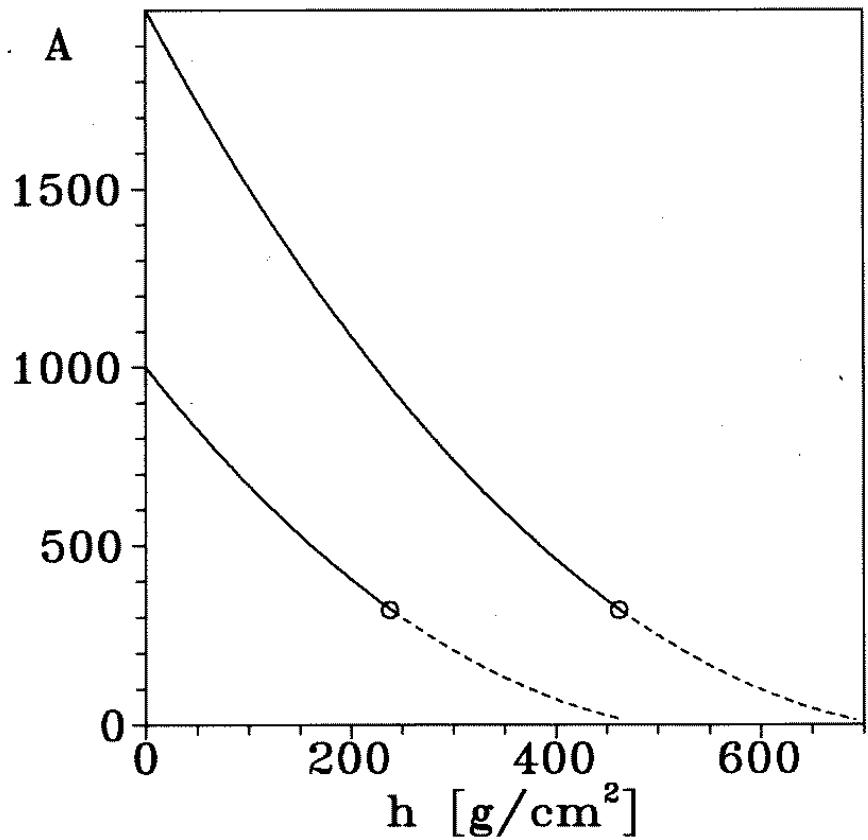}
\end{picture}
\caption{An example of the predicted decrease of the
actual size of the strangelet (as given by its mass number $A$)
with depth $h$ of the atmosphere traversed (measured in g/cm$^2$) for
two different initial SQM mass numbers: $A_0 = 1000~{\rm and}~2000$.
Solid lines corrrespond to $A > A_{crit}$ and dashed to $A <
A_{crit}$ (in which region strangelets practically dissolve into
neutrons).  }
\end{figure}
\newpage
\begin{figure}[h]
\setlength{\unitlength}{1in}
\begin{picture}(4.867,5.000)
\includegraphics{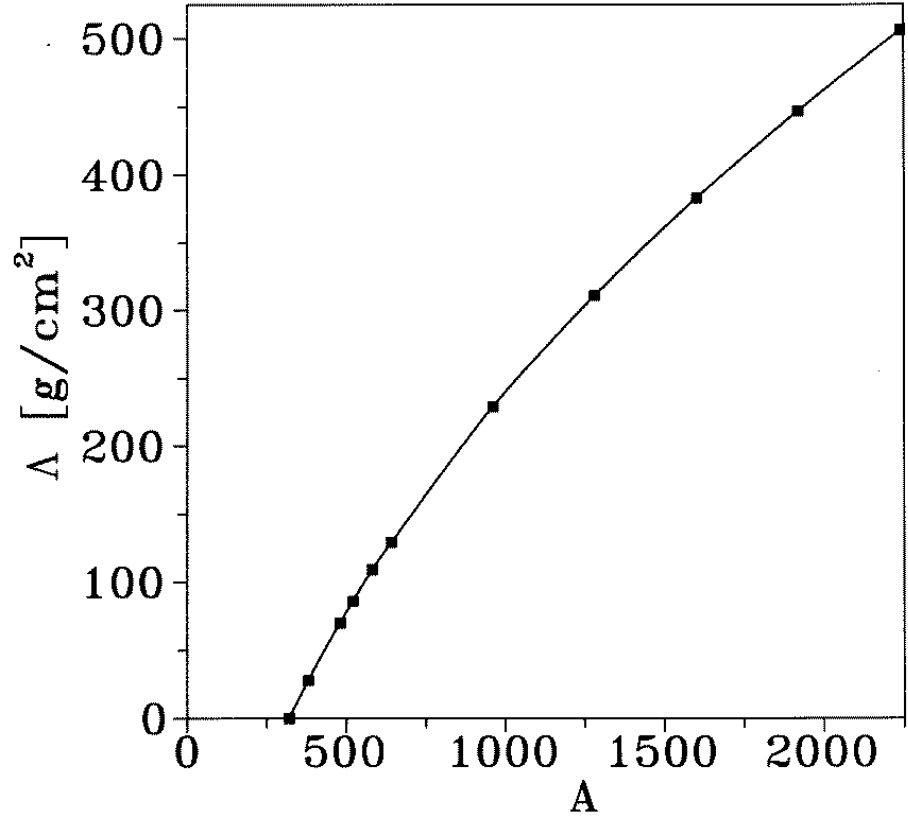}
\end{picture}
\caption{Atmospheric length $\Lambda$ [g/cm$^2$] after
which initial strangelet reaches its critical dimension, $A =
A_{crit}$ drawn as a function of its initial mass number $A_0$
(here $dE/dA > m_N$ for $A_{crit} \leq 320$). Consecutive full squares
indicate (for $A>600$) points where $A/A_{crit} = 2,~3,~4,~5,~6$,
respectively. }
\end{figure}

\end{document}